\documentstyle [12pt,twoside]{article}

\begin{document}
\centerline{\large{\bf Direct Detection of Gravity Waves through}}
\centerline{\large{\bf High-Precision Astrometry}}

\vspace*{1.cm}
\centerline{\bf Redouane Fakir}
\vspace*{0.5cm}
\centerline{Cosmology Group, Department of Physics}
\centerline{University of British Columbia}
\centerline{6224 Agriculture Road, Vancouver, B.C. V6T 1Z1, Canada}
\centerline{fakir@physics.ubc.ca}
\vspace*{0.5cm}
\centerline{\small{\em To appear in the Proceedings of the European Space
Agency Workshop on Astrometry in Space}}
\vspace*{1.cm}
\centerline{\bf Abstract}
\vspace*{0.5cm}

It is generally accepted that a first ever direct detection
of gravity waves would herald a new era in astronomy and in
fundamental physics. Ever since the early sixties, increasingly
larger human and material resources are being invested in
the detection effort. Unfortunately, the gravity wave effects one
has had to exploit so far are extraordinarily small and are usually
very many orders of magnitude smaller than the noise
involved. The detectors that are presently at the most
advanced stage of development hope to register extremely rare,
instantaneous longitudinal shifts that are expected to be orders
of magnitude smaller than one Fermi.
However, it was recently shown that gravity waves can manifest themselves
through much larger effects than previously envisaged. One of these
new effects is the periodic, apparent shift in a star's angular position
due to a foreground gravity wave source.
The comparative largeness of this effect stems from its being proportional
not to the inverse of the gravity wave source's distance to the Earth,
but to the inverse of its distance to the star's line of sight.
In certain optimal but not unrealistic cases, the amplitude of this
effect can reach the critical bar of one
micro-arcsecond, thus raising the prospect that the long awaited
first direct detection of gravity waves could be achieved by a
high precision astrometry space mission such as GAIA.

\clearpage

Perhaps the earliest understood physical effect of
gravity waves is their modulating of proper distances
\cite{einstein1,einstein2,weyl,eddington}.
The first bar detectors \cite{weber}, as well
as recent detection projects such as LIGO
\cite{ligo}
 and VIRGO \cite{virgo}, are based on that effect.

The experimental challenge facing such detection
efforts is daunting. The expected distance modulations
have about the same magnitude as the gravity-wave's
amplitude, which is typically smaller than $10^{-22}$
in the vicinity of the Earth. Thus, these experiments
involve detecting shifts much smaller than one Fermi
in distances of the order of a kilometer.

Not long ago, it was proposed to explore an approach
to gravity-wave detection that is based on accelerations of
null, rather than timelike geodesics \cite{fakir1}.

The simplest illustration of this idea is the shifting
of apparent stellar positions due to an intervening
gravitational pulse \cite{fakir2}. Suppose a supernova
flash hits the Earth, coming from the northern
celestial hemisphere. This is
an indication that a gravitational pulse has also
just whipped passed the Earth, and is now interposed
between us and all the southern celestial hemisphere.
It was calculated that the angular positions of
southern stars would then experience apparent shifts
of the order of the pulse amplitude $h$:
\newline\begin{equation}
|\delta\alpha| \approx {1\over 2} h \sin\alpha  \  \ ,
\end{equation}\newline
where $\alpha$ is the angle of incidence of the light
rays with respect to the gravitational wave front.
In the case of ``pulses with memory''
\cite{kovacs,smarr,braginski,grishchuk}
 such shifts
can be quasi-permanent.

Quantitatively, this version of the effect does nothing
to improve the prospects of gravity-wave detection.
The angular shifts resulting from eq.(1) should be
smaller than $10^{-17}arcsec$, while the precision that
seems achievable today in this context through interferometric astrometry
and very-long-baseline interferometry is about $10^{-6}arcsec$.

However, this effect presents a feature that distinguishes
it qualitatively from most others: In eq.(1), $h$ is not
the amplitude of the waves when they meet the Earth.
Instead, $h$ is the amplitude of the waves when they meet the
stellar photons, which only much later reach the Earth.
This amounts to a prospect of {\em remote probing of
gravity waves}.

Since $h<h_{(Earth)}$ in the case above, this feature can only
worsen the observational situation in this particular
illustration. However, this same feature can dramatically
improve the detectability of the effect, if the configuration
defined by the gravity-wave source, the Earth and the light
source is, in a sense, inverted (Fakir 1994a,b.)
 In the new configurations, as we shall see below,
probing the waves at a distance could mean probing them
in regions of space where $h$ is not smaller, but much larger
than $h_{(Earth)}$.

In the previous illustration, the Earth was placed between
the gravity-wave source and the light source. Consider, now,
a situation where it is the gravity-wave source that is
placed between the Earth and the light source.
Then the photons, during their journey towards the Earth,
would have encountered gravity-wave crests with heights
ranging from $h_{(light\ source)}$ to $h(b)$
to $h_{(Earth)}$, where $b$ is the distance of closest approach
between the photons and the gravity-wave source,
the ``impact parameter.''

The hope, of course, is that the photon will ``remember''
the highest amplitude of gravity waves it sees on its
way to the Earth. If so, the analogue of eq.(1) for
the new configuration would exhibit $h(b)$ on the
right-hand side, rather than the much smaller $h_{(Earth)}$ or
$h_{(light\ source)}$. The whole scheme would then amount
to remote probing of {\em strong} gravity-wave sites.

A priori, there are several reasons to fear that this scheme
would not work. The physics of the photons' encounter
with gravity waves is more involved in this latter case
of spherical wave fronts than in the former plane-wave
case.

For example, one could question whether the deflections acquired
 by a photon during
the ``ingoing'' phase (approaching the gravity-wave source)
are not cancelled by deflections during the outgoing phase.
Fortunately, the calculation shows that this is not the case.

One could also wonder if there would be deflections at all
during the outgoing phase: The gravity-wave crests travel
at the same speed as the photon itself. Now, the photon
is only sensitive to {\em variations} in $h$, and it would
see no such variations if it travels along with the gravity
waves. Nevertheless, in most actual situations, photons
and gravity-wave fronts travel {\em at an angle}. Hence,
in the outgoing phase also, photons
may see changes in $h$ and experience deflections.

Let the gravity-wave mode of interest be described by
\newline\begin{equation}
h = {H\over r}\exp \{i\Omega (r-t+t_{ph})\} \ \ ,
\end{equation}\newline
where $t_{ph}$ determines the wave's phase. $H$ is
a constant that encodes the intrinsic strength of the
source.
Working in a spherical transverse-traceless gauge,
projecting the problem onto a plane containing the
Earth and the light and gravity-wave sources, and considering
the optimal alignment case where $b\Omega$ is of order $1$,
 one finds \cite{fakir3}
\newline\begin{equation}
|\delta\phi|_{optimal} \approx  {3\over 4} \pi \Omega H
= {3\over 2} \pi^{2} | h(r=\Lambda) | \  \ ,
\end{equation}\newline
where $\Lambda$ is one gravitational wavelength. The angle
$\phi$ is close to $0$, $\pi/2$ and $\pi$ at the light source,
the gravity-wave source and the Earth, respectively.

Let us generalize this result to arbitrary values of the
impact parameter $b$. We can infer from eq.(16) of the
above reference that
\newline\[
\delta\phi \approx {H\over b} e^{i\Omega t_{ph}}
\int_{0}^{\pi} d\phi
\exp\left\{ i\Omega b {1+\cos\phi\over \sin\phi} \right\}
\]
\newline\begin{equation}
\times \left[ \sin\phi - {3\over 2} \sin^{3}\phi
+ i\Omega b \left( {\sin^{2}\phi\over 2} - 1 - \cos\phi \right)
\right]
\end{equation}\newline
(This was obtained by comparing the two ends of the
trajectory: $\phi\approx 0$ and $\phi\approx \pi$. One can
show that $\delta\phi = b [ u_{1}(\phi\approx 0) +
 u_{1}(\phi\approx\pi) ]$ , where $u_{1}$ is the fluctuation
of $1/r$.)

Eq.(4) can be rewritten as
\newline\[
\delta\phi \approx {H\over b} e^{i\Omega t_{ph}}
\int_{0}^{\infty}
{4x e^{ib\Omega x}\over (1+x^{2})^{2}}
\]
\newline\begin{equation}
\times \left[ 1 - {6x^{2}\over (1+x^{2})^{2}}
- {ibx^{3}\over 1+x^{2}} \right] dx
\end{equation}\newline
which integrates nicely to the analytical formula
\newline\[
\delta\phi \approx
{1\over 2} H \Omega e^{i\Omega t_{ph}}
\left[  (b\Omega +1) e^{b\Omega} E_{1}(b\Omega)
\right.
\]
\newline\begin{equation}
\left.
+ (b\Omega-1) e^{-b\Omega} E_{1}(-b\Omega)  \right]
\end{equation}\newline
$E_{1}(z)$ is the exponential integral function
\newline\begin{equation}
E_{1}(z) = \int_{1}^{\infty} {e^{-zt}\over t} dt \ , \ \ \
 Re(z)>0 \  \ ,
\end{equation}\newline
extended analytically to the entire complexe plane except $z=0$.
It is straightforward to verify that eq.(5) integrates to eq.(3)
in the limit $b\Omega<<1$.

Thus,
the gravity-wave-induced deflection is equal to
the wave amplitude at only one gravitational wavelength
from the source, times a factor that decreases slightly
faster than $1/b\Omega$.

Besides the future prospects of achieving angular resolutions
of the order of $10^{-7}arcsec$ for radio sources by space-based
 interferometry,
there has been considerable progress, recently, towards reaching
a very high angular precision for optical sources as well
(see \cite{hipparcos,roemer} and several papers in these
proceedings.)
Also, the increase in angular resolution power has been accompanied
by a considerable improvement in photometric sensitivities, potentially
revealing a number of new stellar systems that could
be relevant to this study.

There are several actual astronomical
configurations to which this approach can be applied.
The candidates fall into two classes. In the first,
the gravity-wave source and the light source are aligned
with the Earth by pure chance. They are two unrelated, far apart
celestial objects. Because of the large number of binary stars
in the Galaxy, also because of their relatively large gravity-wave
amplitude and wavelength, a lucky alignment of a binary star
with some more distant light source would be the typical
candidate in this class. Neutron stars are too scarce, are
too weak gravity wave emitters, and the most interesting have too short
wavelengths to qualify for astrometric detection.

Numerically, candidates in this first category could
produce optimal shifts of about $10^{-6}arcsec$, which
falls within the precision attainable by a space-based
astrometric project such as GAIA. Such shifts could be
produced, for instance,
by a very fast binary source with $H \sim 5$cm and an orbital
period of about an hour (i.e. the gravity-wave period is
$2\pi/\Omega \sim 30$ minutes.) For more details about how
GAIA could serve as a gravity wave detector using this effect,
see \cite{makarov}. It is argued there that it could be
possible to detect gravity waves through the effect described
above within an astrometric mission
like GAIA, by scanning the sky systematically for all possible
galactic gravity wave sources, including the (in principle) very
numerous invisible neutron-star binaries.

Alternatively, and perhaps as a first detection attempt that would
be in keeping with the present outline of the GAIA project,
one could first select a not too distant gravity wave
source, typically a fast binary system that is within 100 to 1000
parsecs from the Earth. One then has to find a background star
that lies within a few arcseconds of the binary.
The proper motion of the binary increases
the likelihood of such alignments over a few years period.
There is a number of promising sources in the galaxy to be
investigated for such alignments with background stars, including
cataclysmic variables and massive X-ray binaries
\cite{search}. Because the gravitational signal would be
1) periodic and 2) known to a very high accuracy, it is possible
to use techniques such as data folding and other versions of filter
matching to reduce the noise to the expected level of the gravitational
signal.

A second class of a priori candidates is formed by cases where
the gravity-wave source
and the light source are locked into tight gravitationally
bound systems. Common examples of this in the galaxy are
stars (as light sources) and binaries (as gravity-wave sources)
locked into multiple-star systems or even globular clusters.
Of particular observational importance is the case of
a binary formed by a neutron star (as the gravity-wave source)
and some companion star (as the light source.)

Comparison of typical gravitational wavelengths and
typical separations shows that the
alignment requirement, for this category, is satisfied
naturally.
Unfortunately, because of the
proximity between light source and deflector,
the observationally relevant apparent angular shift of the stellar
image is much smaller than the deflection angle.
Eventually, however, it was shown that another effect could be exploited
in the detection of gravity waves from some of the most
interesting members of this class of candidates, namely those
which comprise a pulsar \cite{fakir3}.

Take, for instance, a system like the well studied binary pulsar
PSR B1913+16 \cite{taylor}. It turns out that this system and
alike could be very promising sites
for direct gravity-wave detection. (This is, of course, besides the
indirect evidence for the existence of gravity waves already provided
by the observed secular slow-down of this binary pulsar.)

Consider first, as the gravity-wave source, the dark neutron
star that revolves around the actual $17$Hz pulsar.
Once every $7^{h}45'$, the two stars come to within only
one light-second (about half a solar radius) of each-other.
This is, at most, of the same order of magnitude as the darker
companion's gravitational wavelength.

In principle, there are two more sources of gravity waves that could
be affecting the apparent position of that same light source.
One is the pulsar itself. Being a neutron star that rotates
$17$ times per second, it should be emitting gravity waves
at a frequency of $34$Hz. However, 1) the angle between the
electromagnetic and the gravitational directions of propagation
is very small in this case, 2) here there is no incoming, only
an outgoing phase. As mentioned above, the combination
of these two facts means that the light from the pulsar is
unlikely to be deflected by the pulsar's own gravity waves.

The other additional source of waves is the binary system as a
whole. (These are the waves for which there is
already indirect observational evidence.) Here also,
the shortness of the incoming phase and the smallness
of the angle between the electromagnetic and the gravitational
directions of propagation are a concern. More importantly,
there are more considerations that have to be taken into account,
before one can make predictions in this case. The photons, here,
originate from the gravity-wave source itself, and traverse
the near-zone ($r<\Lambda$) before reaching the radiation zone,
where our calculations are valid. Such cases necessitate a separate
study, where, in particular, dynamical Neutonian contributions to
the deflection would have to be included.

Numerically, the deflections produced in this case
may well reach the $10^{-6}arcsec$
\cite{fakir3}. This could be the case if,
for instance, the companion is a $10$Hz neutron star
(i.e. the gravitational frequency is $\Omega/2\pi = 20 sec^{-1}$),
radiating perhaps through the Chandrasekhar-Friedman-Schutz mechanism
\cite{chandra,friedman}, with a strength
$H = 10^{-6}$m. However, as we mentioned earlier, these deflections
do not translate into significant apparent angular shifts in this case.
Nevertheless, the consideration of pulsars as sources of the deflected light
lead to another gravity-wave detection prospect, which
we now summarize for completeness.

Following the above study, the next logical step is to try
to exploit the exceptional properties of pulsars, especially
the high stability of their period. This was achieved through
the exploitation of an effect that has little
to do with light deflection,
namely the gravity-wave-induced {\em modulation of time delays},
in the very same astronomical configurations discussed above \cite{fakir4}.

The possibility of gravity-wave detection through the
modulation of pulsar frequencies by {\em plane} waves
has already been extensively explored
\cite{detweiler,romani,hellings,stinebring}.
The experimental effort in this field has made it
possible to detect fractional frequency modulations as faint as
$10^{-15}$ Thus,
stringent upper limits could be imposed on
the cosmological and the galactic
gravity-wave backgrounds. Recently, we also learned that,
in the wake of this effort, the contribution of individual
binary stars was also considered in one instance \cite{sazhin}.
Unfortunately, this initial investigation was not
followed up by the
consideration of more promising candidates, such as systems
consisting of
gravitationally bound light and gravity-wave sources.

In complete analogy with our discussion of the
gravity-wave-induced light deflection effect,
the hope here is that, 1) it can be
shown rigorously that the crossing of a zone of spherical gravity waves
does result in a net frequency modulation;
2) the strongest gravity-wave amplitudes encountered along
the trajectory do contribute to the net modulation.

The same worries we had initially for the working
of the light-deflection effect (see above), can be expressed
here.
Once again, the calculation shows these worries not to
be founded \cite{fakir4}.
 It was shown
that spherical gravity waves can induce time-delay fluctuations
$\delta(\Delta t)$ that vary at a rate
\newline\[
{d\ \over dt_{ph}} \delta(\Delta t)  \approx
{1\over 2} \Omega H  e^{i\Omega t_{ph}} \ \ \ \ \ \ \ \ \ \
\]
\newline\begin{equation}
\times \int_{\phi_{initial}}^{\phi_{final}}d\phi
\sin\phi \exp\left\{ ib\Omega{1+\cos\phi\over \sin\phi} \right\}
\end{equation}\newline
($\Delta t$ is the total time it takes a photon to travel
from the light source to the Earth, via the gravity-wave
source.)

Here also, the problem has a compact analytical solution.
A change of variables can put eq.(8) in the form
\newline\begin{equation}
\left| {d\ \over dt_{ph}} \delta(\Delta t) \right| \approx
2 \Omega H \left|
\int_{0}^{\infty} {x\over (1+x^{2})^{2}} e^{i b \Omega x} dx
\right|
\end{equation}\newline
This integrates to
\newline\[
\left| {d\ \over dt_{ph}} \delta(\Delta t) \right| \approx
H \Omega
\left| {b\Omega\over 2}
\left( e^{b\Omega} E_{1}(b\Omega)
\right.\right.
\]
\newline\begin{equation}
\left.\left.
- e^{-b\Omega} E_{1}(-b\Omega) \right) - 1 \right|
\end{equation}\newline

Hence, numerically, this second effect behaves just like
the first one
(eq.(6)),
at least in orders of magnitude: For optimal alignments, it
is as high as the waves' amplitude only one gravitational
wavelength away from the source. For larger impact parameters,
the effect decreases roughly like $1/b\Omega$.

To use the same numerical illustration as for the previous effect,
a binary star with $H\sim 5$cm and a gravity-wave period (half the
orbital period) $T \sim 2\pi/\Omega\sim 30$ minutes, would produce fractional
frequency modulations of about $5\times 10^{-13}$. A neutron star
with $H \sim 10^{-6}$m and $T = 2\pi/\Omega \sim 0.05$sec, would yield
frequency
modulations as strong as $10^{-12}$.

Retrospectively, this latter approach to gravity-wave detection has exploited
a perhaps curious observational fact. For several cases of gravity-wave
sources that are members of gravitationally bound stellar systems,
the stellar separations can be as small as only one gravitational
wavelength or so. Thus, in a dense globular cluster, the average
stellar separation is of the same order of magnitude as the
gravitational wavelength of a typical binary star. For a binary system,
one member of which is a neutron star, the orbital size can be
comparable to that neutron-star's gravitational wavelength.
Hence, there exists many astronomical sites where light sources are
constantly moving close to, or even within gravity-wave near-zones.

To summarize, what happens in this detection scheme in the most common case
(which is the one
relevant to high precision astrometry) is that
the photons from the background star travel huge
distances virtually unperturbed,
then cross regions of strong gravity waves where their direction
of propagation is shifted, then travel on towards the Earth
where they eventually deliver the record of their encounter with
strong gravity waves. This amounts to a possibility of probing regions
of {\em strong} gravity waves at a distance, thus avoiding the
extraordinary smallness that plagues most other gravity wave effects.

\section*{ACKNOWLEDGMENTS}

I benefited greatly from numerous discussions with W.G. Unruh.
I would also like
to thank U. Bastian, E. H\o g, V. Makarov and H.-J. Tucholke for informative
conversations.

This work was made possible by the sustained
material support of the Cosmology Group in the Departement of Physics,
University of British Columbia.

\end{document}